# Inverse Ising effect and Ising magnetoresistance

Duo Zhao,[1] Jiaqian Sun,[2,3] Wei Tang,[1] and Yu-Jia Zeng[1*]

[1] Key Laboratory of Optoelectronic Devices and Systems of Ministry of Education and Guangdong Province, College of Physics and Optoelectronic Engineering, Shenzhen University, Shenzhen 518060, China

[2] Key Laboratory of Semiconductor Materials Science and Beijing Key Laboratory of Low Dimensional Semiconductor Materials and Devices, Institute of Semiconductors, Chinese Academy of Sciences, Beijing, 100083, China.

[3] Center of Materials Science and Optoelectronics Engineering, University of Chinese Academy of Sciences, Beijing 100190, China.

*Correspondence to: yjzeng@szu.edu.cn

**Abstract**

Ising (Zeeman-type) spin-orbit coupling (SOC) generated by in-plane inverse asymmetry has attracted considerable attention, especially in Ising superconductors and spin-valley coupling physics. However, many unconventional observations and emerging physical phenomena remain to be elucidated. Here, we theoretically study the spin texture of $\sigma_z$ (spin angular momentum projection along $z$) induced by Ising SOC in $1T_d$ WTe$_2$, and propose an unconventional spin-to-charge conversion named inverse Ising effect, in which the directions of the spin current, spin polarization and charge current are not orthogonal. In particular, we predict the Ising magnetoresistance, whose resistance depends on the out-of-plane magnetic momentum in WTe$_2$/ferromagnetic heterostructure. The Ising magnetoresistance is believed to be an interesting counterpart to the well studied spin Hall magnetoresistance. Our predictions provide promising way to spin-momentum locking and spin-charge conversion based on emerging Ising SOC.

**Introduction**

Spin-orbit coupling (SOC) which is induced by crystal's inverse asymmetry, plays a key role in many emerging physical phenomena such as Majorana zero-energy mode, Weyl nodes and so on[1-7]. In particular, Spin Hall effect, spin-orbit torque, and many other magnetoresistances (MR) are also induced by SOC[8-13]. The study of various types of SOC such as Rashba SOC and Dresselhaus-SOC has greatly promoted the development of spintronics[2, 14]. In addition, Ising (Zeeman-type) SOC, which is generated by in-plane (IP) inversion asymmetry of two-dimensional (2D) materials, has attracted much attention, especially in Ising superconductors and spin-valley coupling on collective quantum phenomena[15-19]. In transition metal dichalcogenides (TMDs), especially $1T_d$ WTe$_2$, detailed physical exploration on Ising SOC has been greatly motivated by recent discoveries of out-of-plane damping like torque, free field magnetization switching and unconventional spin-charge conversion [20-25].

SOC can also induce different types of MR, which are also useful tools to explore the spin-momentum locking characteristic[26]. Spin Hall magnetoresistance (SMR) is a good platform to reveal the spin transmission across the heavy metal/ferromagnet (HM/FM) interface [12, 27-29]. Also, the Rashba-Edelstein MR induced by Rashba SOC was observed in Bi/Ag/CoFeB and 2D/ferromagnet heterostructures[30-33]. SMR and Rashba-Edelstein MR have similar performances and therefore are hardly distinguished experimentally. The appearance of SMR can be regarded as the combination of the spin Hall effect (SHE) and inverse spin Hall effect (ISHE)[34]. And the physical mechanism of the Rashba-Edelstein MR can be described as the combination of Rashba-Edelstein effect (inverse spin galvanic effect)[35] and the inverse Rashba-Edelstein effect (spin galvanic effect)[36-38].

As an emerging SOC, the spin polarization direction, charge current and spin current occurring of the Ising SOC are not orthogonal to each other in the processes of spin-to-charge conversion and its inverse effect. Therefore, the MR induced by Ising SOC would have difference performances with SMR or Rashba-Edelstein MR.

In this study, we investigate the spin component along the $z$ direction ($\sigma_z$) induced by Ising SOC in $1T_d$ WTe$_2$ with IP inversion asymmetry. According to the symmetry analysis and first principles calculation, the distribution of the $\sigma_z$ in the band structure is well described. Based on this, we clearly explain that the spin-charge conversion caused by Ising SOC does not obey $\boldsymbol{j}_c \propto \boldsymbol{j}_s \times \boldsymbol{\sigma}$ anymore, where $\boldsymbol{j}_c$, $\boldsymbol{j}_s$ and $\boldsymbol{\sigma}$ are charge current, spin current and spin current polarization, respectively. Most importantly, we predict a new type of MR, namely Ising MR, whose propertis are different from SMR or Rashba-Edelstein MR, i.e., when the magnetic moment (*M*) of ferromagnet points to out-of-plane (OOP), the resistance states are high and low for SMR and Ising MR, respectively.

**Results and discussion**

For $1T_d$ WTe$_2$, there are mirror symmetry along *b*, and the $C_{2a}$ symmetry is broken by the slightly distortion of W-Te bonding as shown in Fig. 1 (a) and Fig. S1 (a) [39, 40]. Then the direction of asymmetry is along to the *b* (*y*) axis (refer to Supplemental Material [41] S1 for details). The

Hamiltonian of SOC can be written as $H_{SOC} \propto bk_x\sigma_z$, which is the expression of the Ising SOC, where $\vec{\sigma}, \vec{p}, \vec{n}$ and $b$ are Pauli matrix, momentum, the unit vector of the asymmetry direction and SOC strength, respectively. The spin texture of Ising SOC can be described by Fig. 1 (c). This means the spin polarization of the electrons in $1T_d$ WTe$_2$ with $k_a$ wave vector would point to $z$. Compared with Rashba SOC $H_{SOC} \propto (k_x\sigma_y - k_x\sigma_y)$ displayed in Fig. 1 (b), the appearance of $\sigma_z$ would broaden our understanding of spintronics, that is when applied external current along to the *a* axis of $1T_d$ WTe$_2$ as shown in Fig. 1 (d), the spin polarization (illustrated in Fig. S1 (b)) along $z$ is generated. Then the spin current ($J_S$) is collinear with the spin polarized orientation. The torque induced by $\sigma_z$ can be expressed as $\boldsymbol{\tau} \propto \boldsymbol{M} \times (\boldsymbol{M} \times \boldsymbol{z})$. This is the out-plane damping like torque found by a series of works recently[20, 21], which is caused by Ising SOC actually. Obviously, the $\boldsymbol{j_c}, \boldsymbol{j_s}$ and $\boldsymbol{\sigma}$ induced by Ising SOC will not obey $\boldsymbol{j_s} \propto \boldsymbol{j_c} \times \boldsymbol{\sigma}$ anymore.

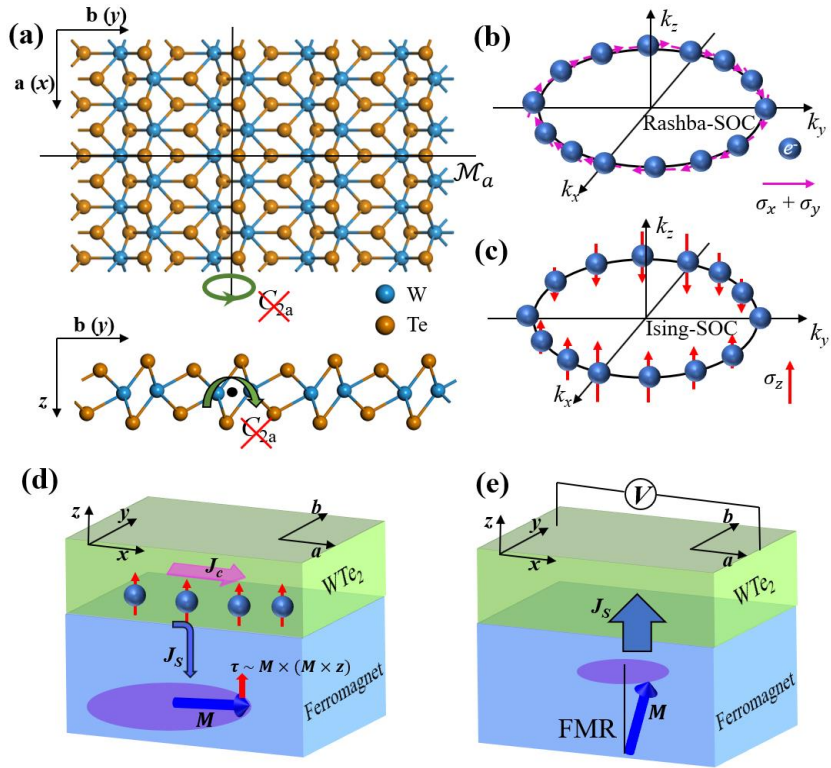

Fig. 1 (a) The top and side views of the crystal lattice $1T_d$WTe$_2$ with given crystal symmetry. (b) and (c) The spin texture of the Rashba SOC (b) and Ising SOC (c). (d) The schematic model of the out of plane damping torque caused by Ising SOC. (e) Illustration of unconventional named inverse Ising effect detected along $x$ direction in $1T_d$ WTe$_2$ by injecting the spin current generated by ferromagnetic resonance (FMR).

As for named the inverse Ising effect (similar with spin galvanic effect) shown in Fig. 1 (e), it is also different from the description of ISHE $j_c \propto j_s \times \sigma$. WTe$_2$ layer would adsorb the spin current generated by ferromagnetic resonance and produce the voltage which is along to the *x* direction under the action of Ising SOC, where $j_s$ and $\sigma$ are parallel with *z*. For details, we will discuss in Fig. 3. At this condition, the generated voltage in the *b* direction would be zero due to the mirror symmetry in the *a* direction as shown in Fig. S1 (c). While only when the spin polarized direction is along to *y*, the voltage measured at *x* direction generated by ISHE would be non-zero as shown in Fig. S1 (d).

The presence of the $\sigma_z$ generated by Ising SOC is analyzed from the symmetry of the crystal structure in Supplemental Material [41] S1. To describe the distribution of $\sigma_z$ in the first Brillouin zone, we performed density functional theory calculations [42] (refer to Supplemental Material [41] S2 for details). The three-dimensional (3D) band structures of $1T_d$ WTe$_2$ is shown in Fig. 2 (a) accompanying with spin projected $S_z$ (marked by grey arrows). The length and directions of the grey arrows represents the one of splitting band's projecting values of the $S_z$ and spin directions, and the standard length of $S_z$=1 is shown at the right side of Fig. 2 (a). The top and bottom color maps represent the energy values of the conduction band and valence band, respectively. To well describe it, it is supposed that $\Gamma$ point is the origin point of the first Brillouin zone, more description of the Brillouin zone and special points are displayed in Fig.S 2 (a). Based on the DFT calculation, the distribution of the $\sigma_z$ is consistent with the symmetry analysis above. Compared with the points (-$k_a$, -$k_b$), the $\sigma_z$ of the points ($k_a$, $k_b$) possess the opposite direction and the same |$S_z$| values, this accord with the characteristic of the $H_{SOC} \propto bk_x\sigma_z$.

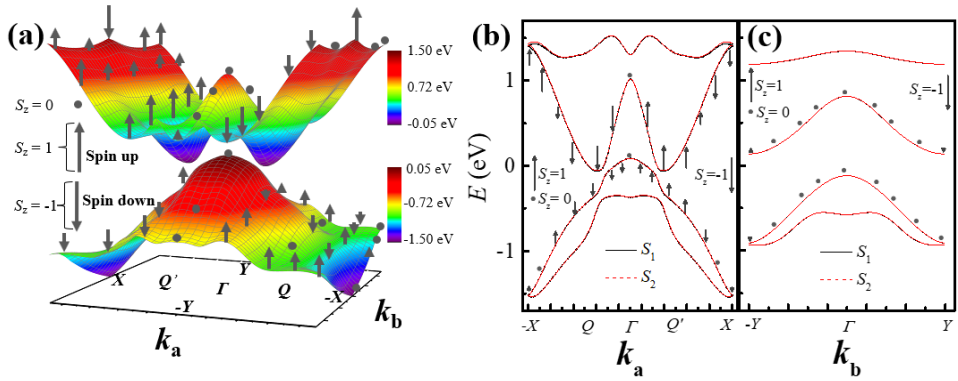

Fig. 2 (a) 3D band structure and spin texture of the $1T_d$WTe$_2$ in the first Brillouin zone consideringSOC. The colors represent the energy values as indicated by color map right side, and the arrow length and direction represent one of splitting band's projected spin values ($S_z$) and spin directions. (b) and (c) The band structure with $\sigma_z$ distribution along $\boldsymbol{k}_a$ ($k_b$=0) and $\boldsymbol{k}_b$ ($k_a$=0) axes extracted from (a), the red dash line and black line represent the spin up and spin down band structure.

To directly display the $\sigma_z$ distribution along $\boldsymbol{k}_a$ ($k_b = 0$) and $\boldsymbol{k}_b$ ($k_a = 0$) axes, the band structures of these two axes are shown in Fig. 2 (b) and (c), respectively. The $\sigma_z$ component caused by Ising SOC at the valence and conduction bands have large values, especially in the conduction band $\boldsymbol{k}_a$ ($k_b = 0$) axis. However, the distribution of $\sigma_z$ along $\boldsymbol{k}_b$ axis shown in Fig 2(c) is almost $S_z = 0$. These DFT results are consistent with the analysis in Fig. 1 (c). There are two energy valleys at $Q$ and $Q$' points, and the values of $S_z$ component are also large[43]. In addition, there are also relatively large band splitting (the splitting of black line and red dashed line) induced by SOC in Fig. S2 (d) comparing with $\Gamma$ point as shown in Fig. S2 (c) obtained from bule and red square enclosed area in Fig. S2 (b). Combining with the appearance of large $\sigma_z$ component, this band spin splitting is mainly induced by strong Ising SOC. Especially, they approach to the Fermi surface, which indicates Ising SOC can result in significant influence on the electrical transport properties of WTe$_2$. Besides, there are also relatively large spin splitting as shown in Fig. S2 (e), which origins from black square enclosed area in Fig. S2 (b). However, this region is far away from the Fermi surface, less influence is expected for the electrical transport properties of WTe$_2$. As for band structure along the $\boldsymbol{k}_b$ ($k_a = 0$) axes, no spin splitting is observed at both conduction and valence bands in Fig. S2 (g) and (h) obtained from zooming in bule and black square enclosed region in Fig. S2 (f). This is also consistent with the characteristic of $\sigma_z$ distribution. Thus, the electrical transport properties of WTe$_2$ are greatly influenced by Ising SOC.

Based on the discussions, the appearance of $\sigma_z$ in 1$T_d$ WTe$_2$ would cause "unconventional" spin-to-charge conversion[44]. To describe these unconventional phenomena concretely, we introduce a spin-to-charge model as shown in Fig. 3 (a)[45]. A charge current is applied to a ferromagnetic electrode (F1) and flows out from electrode F2. The spin polarized current from F1 would pour into the graphene channel. Driven by the spin potential, the spin current would flow toward to 1$T_d$ WTe$_2$ side. Only considering the action of inverse Ising effect (the spin-galvanic effect induced by Ising SOC), the WTe$_2$ layer would adsorb the spin current with spin polarized direction along $z$ and produce the voltage. When the magnetization ($\boldsymbol{M}$) direction of F1 pointed to IP, the spin current without OOP spin polarized component would not produce the spin galvanic effect according to the Ising SOC spin texture shown in Fig. 1 (c) (refer to Supplemental Material [41] S5 and Fig. S4 for details).

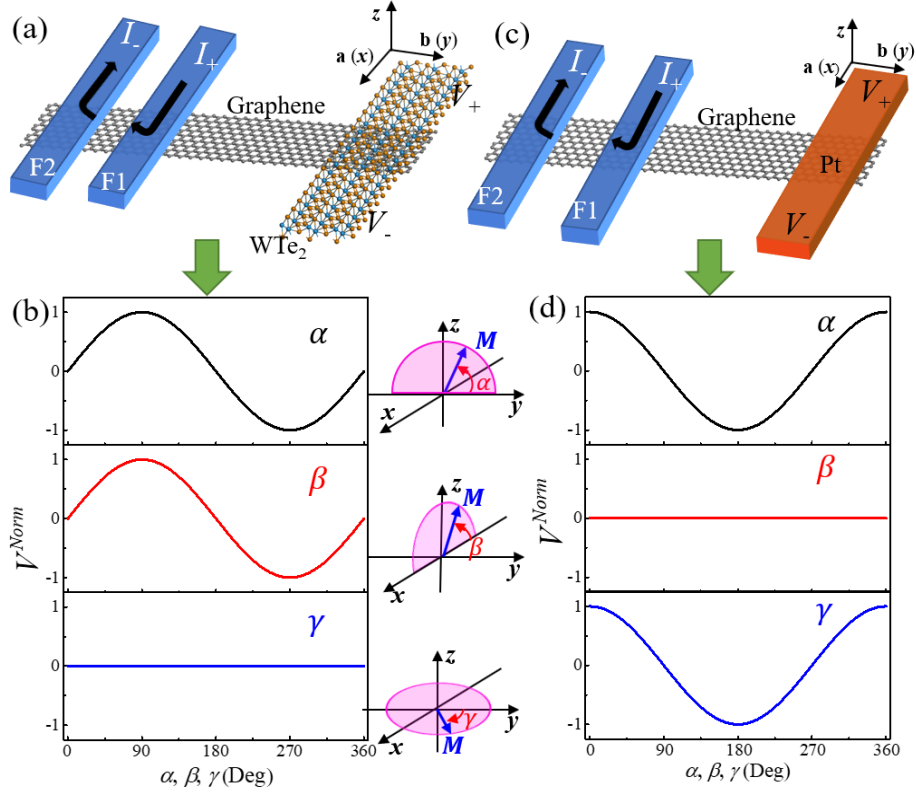

Fig. 3 (a) Sketch structure to confirm the inverse Ising effect induced by $1T_d$WTe$_2$. (b) The change of the normalized voltage ($V^{Norm}$) generated by inverse Ising effect of (a) with $\alpha$, $\beta$ and $\gamma$ under rotation of the $M$ of ferromagnetic electrode (F1) in three types of planes. (c) The schematic model to character inverse spin Hall effect induced by Pt. (d) The $V^{norm}$ vs $\alpha$, $\beta$ and $\gamma$ curves in three different planes generated by inverse spin Hall effect in (c).

Rotating $M$ of the F1 will change the voltage values, and the normalized voltages ($V^{Norm}$) induced by the rotation of $M$ at three different planes are displayed in Fig. 3 (b). When the external magnetic field pulls the $M$ rotating at the $y$-$z$ plane, and the angle between $y$ and $M$ is represented by $\alpha$ as shown by top panel of Fig. 3 (b). Sweeping $\alpha$ from 0°-360°, $V^{Norm}$ is oscillating, which relies on the projection of the spin current $\sigma_z$ component. When $M$ rotated in the $z$-$x$ plane, the change of $V^{Norm}$ with $\beta$ (the angle between -$x$ with $M$) as shown in middle panel of Fig. 3 (b) is same with $\alpha$ due to the same action of projected $\sigma_z$ component of spin current. When $M$ is rotated in the $x$-$y$ plane, the $V^{Norm}$ stays at zero with changing the angle $\gamma$ (the angle between $y$ and $M$) as shown in the bottom panel of Fig. 3 (b). That is because the $\sigma_z$ component of spin current keeps zero during sweeping the *M* of F1 in the *x-y* plane.

When $1T_d$ WTe$_2$ is replaced by Pt as shown in Fig. 3 (c), the changing trend of the $V^{Norm}$ with α, β and γ are different from WTe$_2$ as shown in Fig. 3 (d). The voltage induced by ISHE along *a* direction is mainly contributed by the $\sigma_y$ component of spin current rather than the $\sigma_z$. Therefore, there are marked heavily different performances between Ising SOC of WTe$_2$ and bulk SOC of Pt. In reality, Ising SOC and Rashba SOC may coexist, and the performance of the $V^{Norm}$ may transform from Fig. 3 (b) to Fig. 3 (d) depending on the relative strength of these two types of SOC as shown in Fig. S3 (b). The detailed analysis and explanation can be found in Supplemental Material [41] S3 and S4.

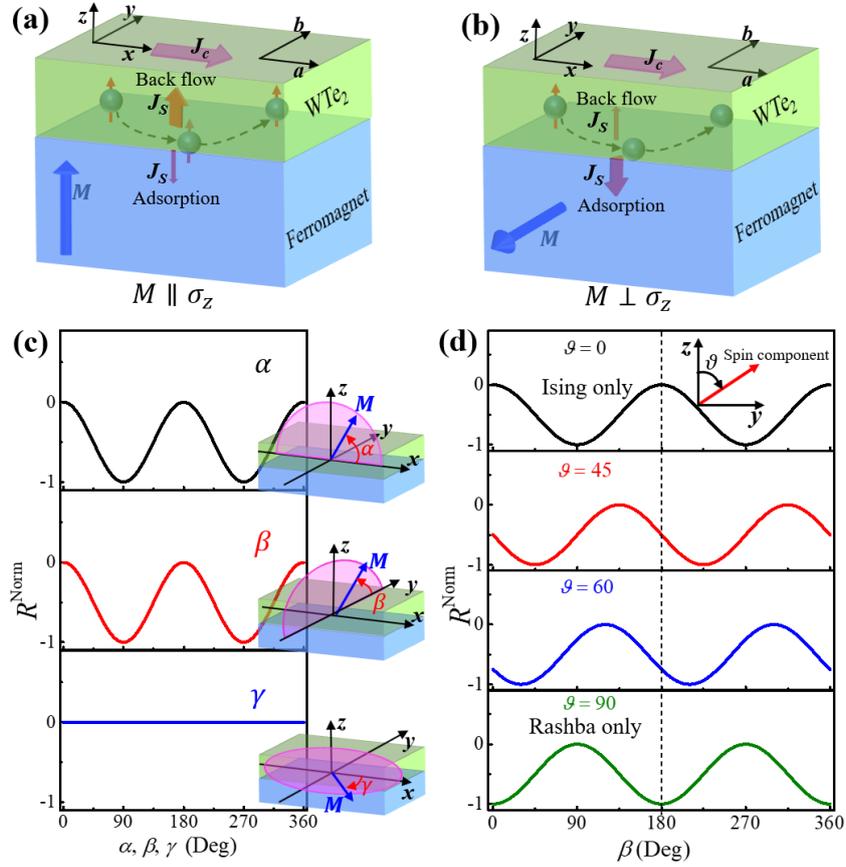

Fig. 4 (a) and (b) The Schematic images of the geometric relation between the flow of electrons and spin accumulation in $1T_d$WTe$_2$/FM heterostructure. (c) The performance normalized Ising MR $R^{Norm}$ vs α, β and γ with rotating *M* of the ferromagnetic layer in three types planes shown in right side. (d) The evolution of the $R^{Norm}$ with spin orbit angle ϑ considering Ising SOC and Rashba SOC in $1T_d$WTe$_2$/FM heterostructure rotation within *y-z* plane.

The $\sigma_z$ spin current can be caused by Ising SOC with unambiguous experimental evidences

[20, 21]. On the other hand, the inverse Ising effect we predict in this study can result in unconventional spin-to-charge conversion, in which the experimental traces were also reported named "unconventional phenomena"[22]. Combining $\sigma_z$ spin current with inverse Ising effect, a new type of MR, namely Ising MR, is predicted here, which has different performance compared with SMR or Rashba-Edelstein MR. The detailed discussion about physical mechanism and signal strength can be found in Supplemental Material [41] S5. In WTe$_2$/FM heterostructure as shown in Fig. 4 (a), an external current $J_C$ flowing through WTe$_2$ layer will generate spin polarization at interface of heterostructure under action of the Ising SOC. Then spin current $J_S$ towards to the FM layer would occur. When the direction of $M$ points to OOP, the $J_S$ with $\sigma_z$ is rarely absorbed by the FM layer, the back flowed $J_S$ will generate the electric field (voltage) due to the inverse Ising effect. At this condition, the measured resistance ($R_{xx}$) is small (without considering anisotropic magnetoresistance). When $M$ of the F layer points to IP as shown in Fig. 4 (b), the spin current with OOP spin polarization is adsorbed by the FM layer and driven the procession of $M$. Then the $R_{xx}$ of WTe$_2$ would stay unchanged.

When an external magnetic field pulls $M$ from OOP to IP, the $R_{xx}$ increases to maximum. To well describe the Ising MR, we discuss the change of $R^{Norm} = \frac{R_{xx} - R^{\perp}}{R^{\perp} - R^{\parallel}}$ at three different planes as shown in Fig. 4 (c), where $R^{\perp}$ and $R^{\parallel}$ represent the resistances when the spin polarized direction of $J_S$ is perpendicular and collinear to $M$, respectively. When the $M$ is rotated in the $x$-$z$ plane, $\alpha$ increases from 0° to 90° as shown in the top panel of Fig. 4 (c). $R^{Norm}$ would decrease from 0 to -1 due to the change of spin polarized direction of $J_S$ with $M$ from orthogonality to colinear state. Increasing $\alpha$ from 90° to 180°, $R^{Norm}$ would return to 0 due to the change of $\sigma_z$ and $M$ from colinear state to orthogonality. Continually sweeping $\alpha$ from 180° to 360°, $R^{Norm}$ would decrease from 0 to -1 and return to 0 again. That is because the change of spin polarized direction of $J_S$ with $M$ would experience from orthogonality to colinear state and then gradually return to orthogonality again. When the $M$ of the F layer is rotated in the $y$-$z$ plane as shown in middle panel of Fig. 4 (c), the change of $R^{Norm}$ with $\beta$ (the angle between $M$ and $y$) is same with $R^{Norm}$-$\alpha$ curve because of their same variation of spin polarized direction of $J_S$ with $M$. As for rotating $M$ in the $x$-$y$ plane, the direction of $M$ and spin polarized direction induced by Ising SOC are always orthogonal, then the $R^{Norm}$ stays at 0 valve and remains unchanged with the variation of $\gamma$ (the angle between $y$ and $M$) as shown in bottom panel of Fig. 4 (c). Compared with SMR shown in Fig. S3 (c) and (d), the performances of the Ising MR are totally different. The Ising MR shown in the middle panel of Fig. 4 (d) has a 90° phase difference compared with SMR as shown in the middle panel of Fig. S3 (d), then the Ising MR can be distinguished from SMR or Rashba-Edelstein MR. Ising MR is therefore a good platform to study the fundamental of spin orbit torque, spin current transport and so on.

In real situation, the Ising SOC and Rashba SOC will coexist inevitably, which influence the performance of $R^{Norm}$. As shown in insert of Fig. 4 (d), $\vartheta$ is the spin orbit angle and the strength of Ising SOC and Rashba SOC can be written as $a = \gamma cos\vartheta$ and $b = \gamma sin\vartheta$ [46] (refer to Supplemental Material [41] S3 for details). The values of the $\vartheta$ represent the relative strength of Ising SOC and Rashba SOC. When $\vartheta = 0°$, it means only Ising SOC exists. As $\vartheta = 90°$, only Rashba SOC exists. The influence of the $\vartheta$ on $R^{Norm}$ vs $\beta$ curves is shown in Fig. 4 (d). The function of $R^{Norm}$ can be written phenomenologically as

$$R^{Norm} = \frac{1}{2}(\cos(2\beta + 2\vartheta) - 1) \qquad (1)$$

As for $\vartheta = 0°$ and $\vartheta = 90°$, the changes of the $R^{Norm}$ with $\beta$ at *y-z* plane are shown in Fig (d) with black and green curves, respectively. When $\vartheta = 45°$, $R^{Norm} - \beta$ curve is displayed in Fig. 4 (d) with red line according to equation (1). At this point, the angle of the spin polarization direction with *z* axis is same with $\vartheta$ as shown by insert of Fig. 4 (d). For $\beta = 45°$ and $225°$, the *M* is collinear with spin polarized direction, and $R^{Norm}$ is minimum. For $\beta = 135°$ and $315°$, the *M* is perpendicular to the spin polarized direction, and $R^{Norm}$ is maximum. When $\vartheta = 60°$, $R^{Norm} - \beta$ curve is displayed in Fig. 4 (d) with blue line. The *M* is collineation with spin polarized direction, and $R^{Norm}$ is minimum at $\beta = 30°$ and $210°$. When $\beta = 120°$ and $300°$, the *M* were perpendicular to spin polarized direction, and $R^{Norm}$ is maximum. Therefore, the relative strength of Ising SOC and Rashba SOC can be separated according to the spin orbit angle $\vartheta$. The value of the $\vartheta$ can be easily obtained from $R^{Norm} - \beta$ curve experimentally.

It is worth mentioning that recently observed unconventional spin-momentum locking phenomenon by "AMR" in single layer WTe$_2$[47, 48] can be well explained by our proposed spin orbit angle $\vartheta$ caused by Ising SOC and Rashba SOC. The angle values may vary for different results, because the Rashba SOC strength depends on interfaces properties of WTe$_2$. However, the Hanle MR[49] may be mixed in the reported "AMR" signals, which could also be induced by Ising SOC and requires further experimental investigations. Also note that the $\sigma_z$ and unconventional spin-to-charge conversion discussed here are different from the observations in non-collinear antiferromagnetic materials[50, 51], in which the spin-to-charge conventions are induced by chiral arrangement of magnetic moments in real space rather than SOC[52].

**Conclusion**

We have demonstrated the generation of $\sigma_z$ considering Ising SOC in $1T_d$ WTe$_2$ from symmetry analysis and DFT calculation, in which the $\sigma_z$ does not obey the rule of $\boldsymbol{j_c} \propto \boldsymbol{j_s} \times \boldsymbol{\sigma}$ as in the case of 2D materials with in plane asymmetry. We predict the existence of so-called inverse Ising effect, which can induce unconventional spin-to-charge conversion. Most importantly, we predict a new phenomenon, namely Ising MR, which originates from the combination of the Ising effect and inverse Ising effect. Ising MR is believed to be a promising complement to the well-known SMR and Rashba-Edelstein MR from both scientific and technical points of view. Our work broadens the understanding of SOC and the related emerging physical phenomena, which provides unprecedented opportunity to the fields of spintronics and multiferroics.

**Acknowledgments**

This work was supported by the National Natural Science Foundation of China (Grant No. 52273298), the Guangdong Basic and Applied Basic Research Foundation (Grant No. 2022A1515010649), and the Shenzhen Science and Technology Foundation (Grant Nos. JCYJ20210324095611032, JCYJ20220818100204010 and JCYJ20220818100008016)